# The need for Professional-Amateur collaborations in studies of Jupiter and Saturn

Emmanuel Kardasis (1), John H. Rogers (2,3), Glenn Orton (4), Marc Delcroix (5), Apostolos Christou (6), Mike Foulkes (2), Padma Yanamandra-Fisher (7), Michel Jacquesson (3), Grigoris Maravelias (1,8)

*(1) Hellenic Amateur Astronomy Association (Athens, Greece),*
*(2) British Astronomical Association (London, UK), (3) JUPOS team,*
*(4) Jet Propulsion Laboratory, California Institute of Technology (Pasadena, CA, USA),*
*(5) Societe Astronomique de France (Paris, France), (6) Armagh Observatory (Armagh, UK),*
*(7) Space Science Institute (Rancho Cucomonga, CA, USA),*
*(8) Astronomický ústav, Akademie ved Ceské Republiky (Ondrejov, Czech Republic)*

## Abstract

The observation of gaseous giant planets is of high scientific interest. Although they have been the targets of several spacecraft missions, there still remains a need for continuous ground-based observations. As their atmospheres present fast dynamic environments on various time scales, the availability of time at professional telescopes is neither uniform nor of sufficient duration to assess temporal changes. On the other hand, numerous amateurs with small telescopes (with typical apertures of 15-40 cm) and modern hardware and software equipment can monitor these changes daily (within the 360-900nm wavelength range). Amateur observers are able to trace the structure and the evolution of atmospheric features, such as major planetary-scale disturbances, vortices, and storms. Their observations provide a continuous record and it is not uncommon to trigger professional observations in cases of important events, such as sudden onset of global changes, storms and celestial impacts. For example, the continuous amateur monitoring has led to the discovery of fireballs in Jupiter's atmosphere, which provide information not only on Jupiter's gravitational influence but also on the properties and populations of the impactors. Photometric monitoring of stellar occultations by the planets can reveal spatial/temporal variability in their atmospheric structure.

Therefore, co-ordination and communication between professionals and amateurs is important. We present examples of such collaborations that: (i) engage systematic multi-wavelength observations and databases, (ii) examine the variability of cloud features over timescales from days to decades, (iii) provide, by ground-based professional and amateur observations, the necessary spatial and temporal resolution of features that will be studied by the interplanetary mission *Juno*, (iv) investigate video observations of Jupiter to identify impacts of small objects, (v) carry out stellar-occultation campaigns.

## 1. Introduction

Collaboration between amateur and professional astronomers is increasingly important in many areas of solar system science [1],[2]. In this paper we focus on the gaseous giant planets. For both professionals and amateurs, the overarching goal of observations is to understand the atmospheres of the giant planets, which are so different from Earth, and so typical of many planets now being discovered around other stars.

Jupiter and Saturn have been observed intensively for over a century. There are historic accounts of these observations in publications, most of which are by British Astronomical Association (BAA) [3], Association of Lunar and Planetary Observers (ALPO) [4], Association of Lunar and Planetary Observers of Japan (ALPO Japan) [5] and Societe Astronomique de France (SAF) [6]. In recent years amateurs have been capturing numerous high-resolution images as a result of the advancement of technology, providing near-continuous coverage in visible and near-infrared wave-bands. Meanwhile, professional astronomers have long-term observational programmes to observe the giant planets with

NASA's InfraRed Telescope Facility (IRTF) and the Pic du Midi 1-m telescope. Observations are also made using the *Hubble Space Telescope (HST)*, Subaru telescope of National Astronomical Observatory of Japan and the European Southern Observatory's Very Large Telescope (VLT). However, most of these are directed to the giant planets only intermittently (largely due to intense competition for time) and mainly at infrared wavelengths. Moreover, several spacecraft have visited each planet, and the *Cassini* spacecraft is operating in orbit around Saturn as of this writing.

Even though the professional ground-based and spacecraft observations provide a large variety of spectral and temporal data, the amateur network of more than 100 observers around the world still provides professionals with important types of information: complementary wavelength coverage to professional near- and mid-infrared observations; global coverage and context for their observations; independent verification and confirmation of their observations; and tracking of features and alerts of new events which are worthy of professional study. Therefore, the key advantage that the amateur community provides is that of flexibility and spatiotemporal coverage for context. Both of these widen the discovery space significantly compared to the capabilities of professional observers.

The paper is structured as following: Section 2 presents professional - amateur (PRO-AM) collaborations currently under way regarding Jupiter and Saturn. Information on the digital observations methodology is provided in Section 2.1 along with the existent databases where they are stored. In Section 2.2 PRO-AM collaborations and results regarding the atmospheric features of Jupiter and Saturn are discussed, while in Section 2.3 the support of space missions by amateur imaging is highlighted. The investigation of impacts on Jupiter's atmosphere is presented in Section 2.4. Stellar occultations as atmospheric probes are discussed in Section 2.5. Section 3 summarizes and concludes the current paper.

## 2. Fields of PRO-AM collaboration

### 2.1 Systematic multi-wavelength observations and databases

Amateurs observe Jupiter and Saturn with a variety of telescopes, mostly with apertures 15-40 cm. Since the early years of the 21$^{st}$ century, amateur work has mostly consisted of imaging with webcams combined with processing by software to select and combine the sharpest images. This technique has been referred to as 'lucky imaging' [1] but we do not use this term as good results require technology, skill, and perseverance, just as much as luck. In recent years cameras based on the advantages of webcam technology have been designed for digital planetary observations (planetary cameras). These cameras capture videos over several minutes, producing hundreds of images in a rate of 15 to 200 frames per second. Software such as Registax [7] or Autostakkert [8] is then used to select the images with the highest-spatial-frequency components, to align and stack them. Further image processing (e.g. applying wavelets, adjusting brightness-contrast) is then performed in these or other photo-processing programs. Normally the images have to be taken within a span of ~2-3 minutes for Jupiter (~4 minutes for Saturn), to avoid smearing by the planet's rotation. However, a recent innovation in the WinJUPOS software suite allows either raw images or stacked images to be 'derotated' to compensate for the planet's rotation, allowing much longer integrations which produce substantially sharper images, especially in mediocre seeing. All this image processing does not preserve absolute intensity values, but it allows a great deal of fine detail to be resolved, and independent imaging and processing by different observers confirms that relative intensities and colours can be realistically, if qualitatively, portrayed.

Images can be taken in monochrome or in colour. While some observers use colour planetary cameras, most use monochrome planetary cameras with a series of colour filters. The most common filters used by amateurs are the red (R ~390-500 nm), green (G ~500-580 nm) and blue (B ~580-670 nm) filters supplied by various companies, the synthesis of which results in colour images. (However, at present amateurs do not have any way of recording absolute colours; colour balance in images is arbitrary, and chosen for best contrast rather than for reproducing the true predominantly yellowish colours of Jupiter and Saturn). Additionally, near-infrared (IR) filters are used (~700-900nm), providing continuum images which penetrate slightly deeper into the clouds, and most importantly the narrow-band $CH_4$ filter (890nm, with typical Full Width at Half Maximum of 5-18 nm depending on manufacturer) that shows light reflected from the highest clouds. A near-ultraviolet (UV) filter may also be used (<390 or <360 nm) (Fig. 1). For Saturn, the >610nm band-pass filter (far red plus near-IR continuum) is very useful for detecting discrete atmospheric spots.

The need for international databases that host this vast number of worldwide observations is fulfilled primarily in two sites: the professional International Outer Planet Watch - Planetary Virtual Observatory and Laboratory database (IOPW - PVOL database) [9] and the amateur ALPO-Japan database [5]. A dedicated database of the position of features from historical and current observations is provided by the JUPOS project (Database for Object Positions on Jupiter) which can be accessed at [10]. Moreover, WinJUPOS [10] is an important planetary software suite developed by Grischa Hahn, enabling amateurs to perform sophisticated analysis of images. Among the various applications it enables construction of global maps, measurement of positions of cloud features, their analysis in drift charts, and examination of their motion in time. Such drift charts are becoming increasingly important to guide professional observations, whether from observatories or from spacecraft, as we discuss below.

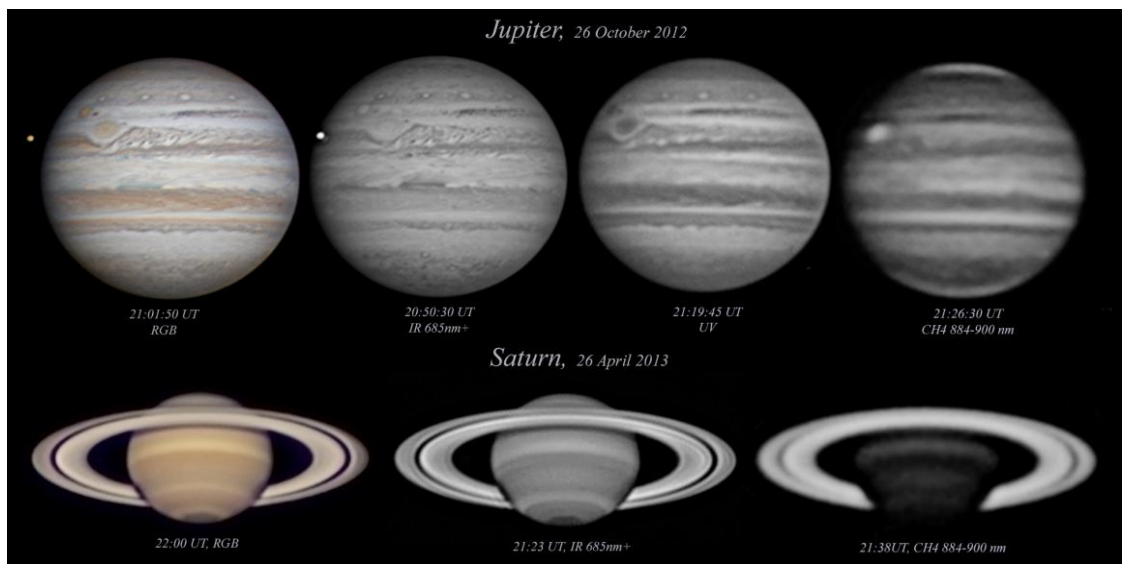

*Figure 1. A set of images showing the different aspects of Jupiter (top) and Saturn (bottom) according to wavelength by co-author E. Kardasis (South is up).*

## 2.2 Examination of the variability of cloud features

The ever-changing atmospheres of Jupiter and Saturn with a large number of atmospheric features (especially Jupiter) need continuous long-term coverage. Moreover, the fast rotation of Jupiter (~9h55m) and Saturn (~10h39m) requires a broad geographic distribution of observers around the world to maximize the coverage of the events. Large-scale climatic cycles occur at intervals of several years (Fig. 2). The onset of these cycles is usually unpredictable, and when they do begin, time availability on professional telescopes is not enough to follow them, thus only amateurs can offer continuous observations over many years. On the other hand, only professionals can observe in mid-IR to detect features below the visible cloud-tops, to do precise high-resolution photometry to retrieve cloud structures, and (using spacecraft missions) to resolve cloud textures sufficiently to provide accurate wind profiles over the whole planet. Atmospheric scientists aim for a synthesis of all these observations, complemented by theoretical and computational modelling, to understand the atmospheres of these planets.

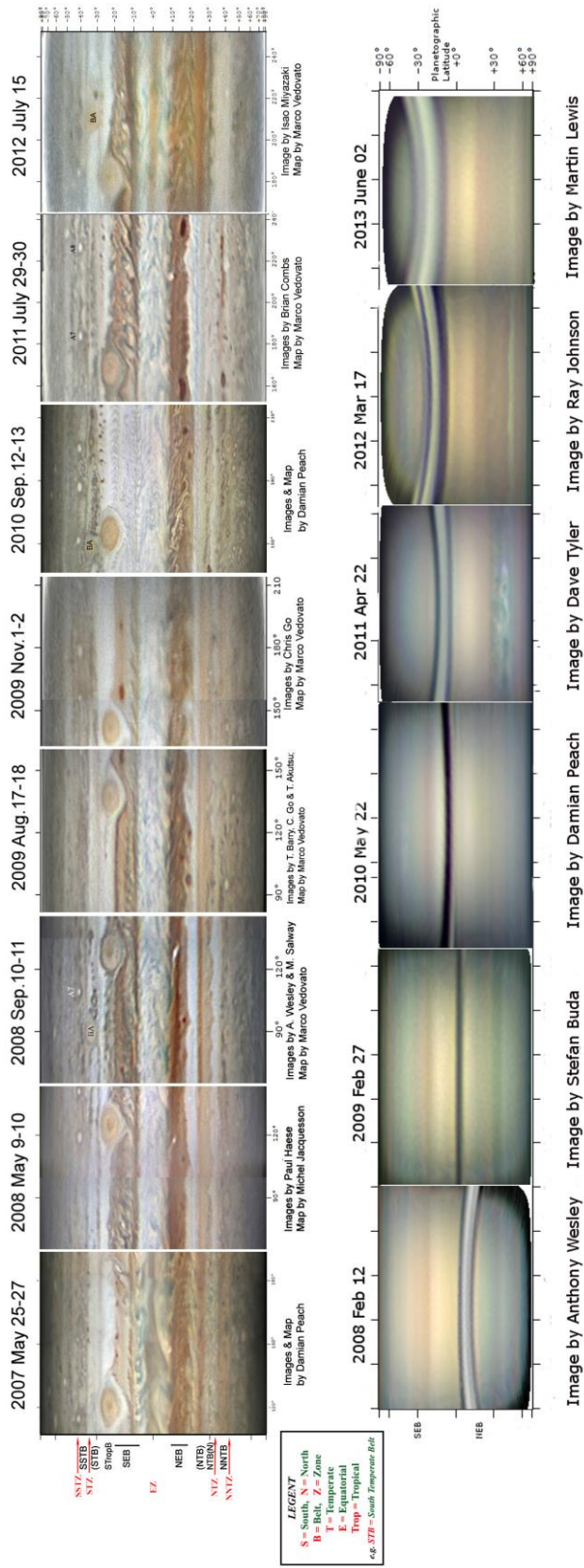

**Figure 2. Top:** *Changes in Jupiter's belts, 2007-2012, in visible colour maps. Cylindrical projection maps made in WinJUPOS are aligned to show the changing latitudes of the edges of the belts. Abbreviations for the major belts are labelled at left. The figure depicts well the major changes in belts and zones from year to year, including South Equatorial Belt (SEB) fading and revival in 2009-2011, and North Equatorial Belt (NEB) narrowing and revival in 2011-2012 [11].* **Bottom :** *Changes in Saturn's belts, from 2009 to 2013 are shown in these cylindrical projection maps made in WinJUPOS and derived from color images (Differences between images in overall colour balance are largely due to different processing by individual observers). These maps are aligned to show the changes in latitudes of the edges of the belts plus major changes in the belts over this period. In particular, the maps show the fading of the NEB during 2010 and the appearance of the Great White Storm in the northern hemisphere during 2010/2011. The maps also show the varying aspect of projection of both the rings and the ring shadow onto the planet, resulting from the varying inclination of the rings with respect to the Earth. (South is up for both planets).*
*Map by co-author M. Foulkes.*

## 2.2.1 Evolution of the atmosphere of Jupiter

### 2.2.1.1 Amateur studies

For Jupiter, more than any other planet, there has been a long history of important amateur research, thanks to the fortuitous fact that most of the planet's major atmospheric features are clearly visible and resolved in moderate-sized amateur telescopes. It was amateurs in the 19th and early 20th century who identified all the regular currents governing the motions of large weather systems in different latitudes; who discovered several of the major jets; and who described recurrent climatic cycles such as the 'fading' and 'revival' of the SEB. Amateurs continued to monitor these features throughout all subsequent years, publishing a continuous record of the visible atmospheric features (summarised in [12] & [13]). This work has continued to the present day, now amplified by the use of modern imaging techniques, with a worldwide network of proficient observers, and modern analysis techniques developed by the JUPOS project [10].

JUPOS is a project of an amateur team in Europe, led by Hans-Jörg Mettig, that uses amateur observations and the WinJUPOS software to track the evolution of Jupiter. Measurements of the positions (longitude/latitude) of all features clearly visible on the image are made, and recorded in a database. The database is first used to produce drift charts, which display the movements of the spots against time for a given latitude span (Fig. 3) or a particular object (Fig. 4A). Second, the graphs allow the computation of the drift rate of each spot against a rotation system, and changes of speed in relation to morphology can be tracked. It is now possible even to calculate the global zonal wind profile [14,15,16].

Finally, using WinJUPOS, images obtained in a time span of no more than 2-3 days can produce a map of Jupiter showing the global aspect of the planet on a given date (Fig. 5). A longer time span would give an inconsistent image of the planet because of: (i) the difference of rotation periods between equatorial and other latitudes, and (ii) the rapid evolution of some features in very active regions.

The results are used by the JUPOS team with co-author J.H. Rogers to make systematic reports of the planet's atmosphere [17]. These provide detailed descriptions of the phenomena with compilations of world-wide amateur images, in order to maintain a continuous record of atmospheric evolution, report new phenomena that are worthy of professional investigation, provide predictions of the locations of features, and estimate the likely outcomes of ongoing disturbances.

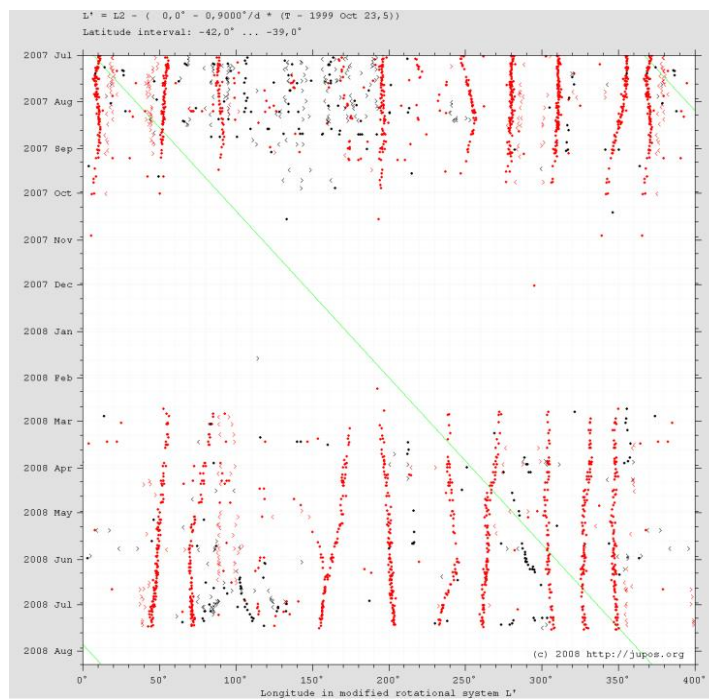

***Figure 3.*** *Tracking of white ovals (shown as red dots) and dark spots (shown as dark dots) in SSTB/SSTZ (see Fig. 2) in 2007-2008 by the JUPOS team. Every dot is a position measurement of a feature. Dense and continuous observations are required to track in detail the motion and drift rate (speed) of meteorological formations. In this chart, the longitude scale is chosen to move at exactly 0.9 deg/day relative to System II, thus approximately matching the mean drift of the long-lived features. The green line is the longitude 0° of System II which is oblique as the graph is made with a modified rotation system L' defined in the header of the graph. JUPOS can also produce charts in System III, but System III is not used routinely because it is repeatedly changed by the International Astronomical Union, whereas System II has been precisely defined for over 100 years and approximately matches the mean drift of the Great Red Spot.*

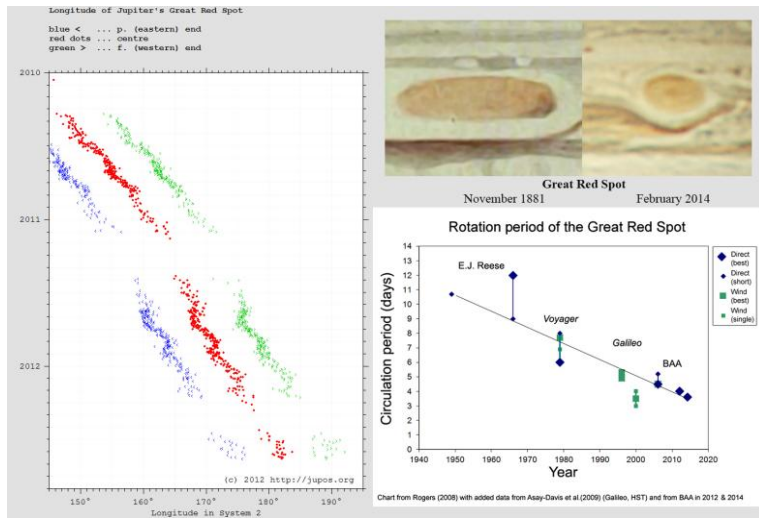

*Figure 4.* Properties of the Great Red Spot (GRS), the largest vortex in Jupiter, which has a variable reddish colour. Amateur contributions to the study of this vortex have been extensive. **A) Left panel:** A drift chart of the Great Red Spot (GRS) by JUPOS team. In this graph we can see part of its long-term historical drift rate and length variation, together with its 90-day zonal oscillation. **B) Top right panel:** Comparison of the size and color of Jupiter's Great Red Spot as drawn by Thomas Gwyn Elger on November 28, 1881 (BAA archives) and as imaged by co-author E.Kardasis 133 years later, on February 27, 2014, (South is up). The length was 34 deg. (39600 km) in 1881 and 13.7 deg (16000 km) in 2014. **C) Bottom right panel:** Chart showing the shortening of the GRS internal rotation period from spacecraft and ground based amateur observations, by co-author J.H. Rogers [18]. This has been updated to January 2014, when amateur observations showed that the GRS was smaller than ever before and its internal rotation period had also shortened [17] Acting on this information, a professional team obtained images with HST at short notice to investigate the changing dynamics of the GRS [19].

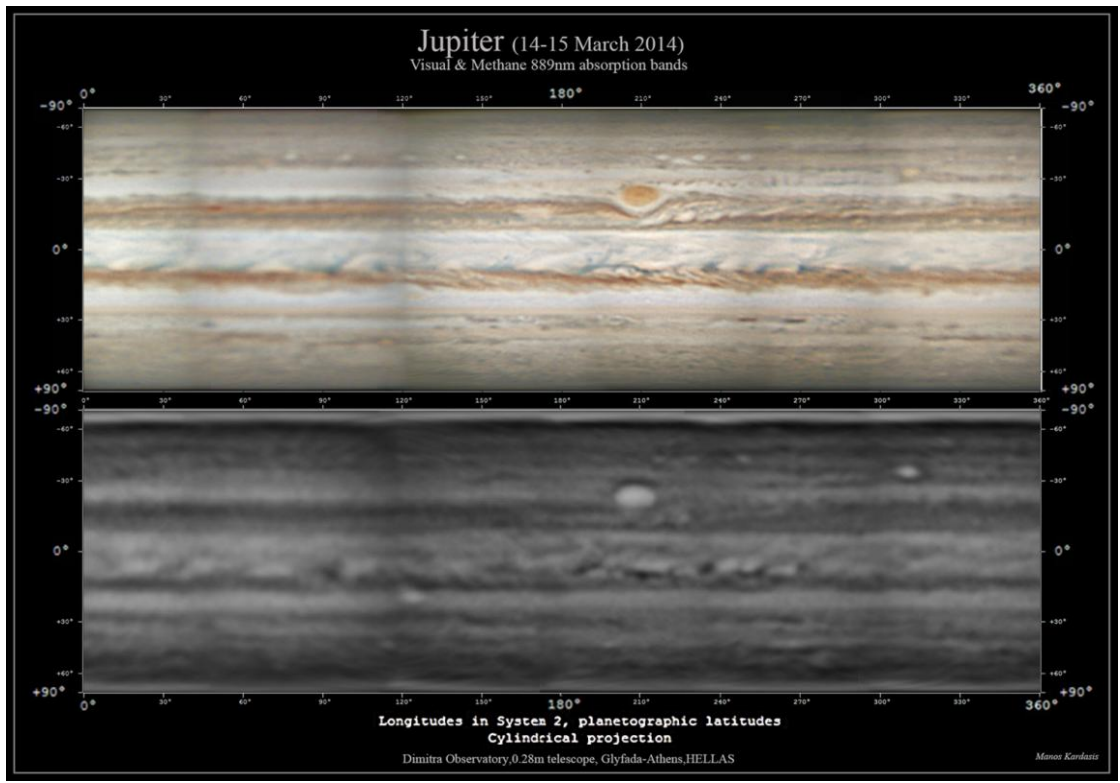

*Figure 5.* Map of Jupiter during 2014 March.14-15. It is made from individual observations of the planet with the use of WinJupos software. **Top panel:** Map at visual wavelengths with RGB filters. **Bottom panel:** Map made with a methane $CH_4$ 890nm filter which reveals information on cloud altitude (brighter areas are higher-altitude formations). Observations near opposition, by a single dedicated amateur can cover a wide swath of longitudes to create a map. (Images and map by co-author E. Kardasis,[20], South is up).

**2.2.1.2 PRO-AM collaborations**

At present, the relationship between amateur work and professional science can be summarized in three categories:

1) Amateur contributions in planetary imaging and analysis:
As presented already in Introduction there is a number of available databases which archive amateur images [3-6,9]. These have served as repositories of valuable data which are used by both the amateur and professional communities to investigate atmospheric phenomena, leading to important PRO-AM collaborations and peer-reviewed publications (e.g. the GRS shortening presented in Fig. 4 [19] or the SEB revival cycle [21]; more are discussed in the next sections). The Planetary Sciences Group UPV/EHU [22] led by Agustin Sanchez-Lavega have published works based largely on amateur images since about 1990, with leading amateurs as co-authors [e.g. 16, 23, 24]. They continue to publish many papers that combine feature tracking from images with physical analysis and dynamical modeling. The group also operates the IOPW amateur network database [9] for which the software tool PVOL has been developed. PVOL helps in finding and selecting the planetary observations taken by amateur astronomers.

2) Alerts:
Amateurs can alert professionals to new atmospheric features or events, or predict the positions of features of interest. These alerts sometimes prompt professionals to target infrared telescopes or spacecraft (*Galileo*, *HST*, *New Horizons*).

3) Complementary support to space missions:
Amateurs can provide information at visible wavelengths to supplement professional observations at infrared wavelengths [25], or 'snapshots' taken by spacecraft. The complementarity of these data sets was well shown during the *Cassini* flyby of Jupiter in 2000/01 [26] and subsequently. Thus, amateur reports can provide context on timescales of days (for matching actual features), months/years (for describing a climatic cycle that may be under way), and years/decades (for understanding the historical pattern of such processes).

**2.2.1.3 Jovian atmospheric features subject to PRO-AM collaborations**

Phenomena of joint interest include the permanent jets, major planetary-scale disturbances, vortices, waves, other disturbances, and ephemeral events such as impacts. The jets constitute the fundamental framework of Jupiter's atmosphere, but until recently only spacecraft were capable of observing their full detail. Nowadays, amateur observations can sometimes reveal the peak speeds of the major jets, and previously unknown variations within them (North Equatorial Belt-NEB [14], North Temperate Belt south-NTBs [27], South Equatorial Belt north-SEBn [28]).

Wave phenomena on the major equatorial jets are of particular interest and recent papers have been able to combine amateur and professional observations and analysis to produce informative synthesis which would not have been possible with either data set alone, for example on thermal waves on the NEB [26] and dynamical waves in the SEBn jet [28, 29].

The most impressive transient phenomena include wholesale fading or narrowing and subsequent violent revivals of the South Equatorial Belt (SEB), North Temperate Belt (NTB), and North Equatorial Belt (NEB) [13, 22, 30, 31]. After many years with no such disturbances, these have recently resumed, with SEB revivals in 2007 and 2010, NTB revivals in 2007 and 2012, and a NEB revival in 2012 (all reported in detail on the BAA web site [17; see also 31]; see also Figs. 2 & 6). Early PRO-AM studies of some of these have been published [24,32,33]. Historically these disturbances have appeared to be dissimilar, with the violent disturbances being initiated by appearance of an intense local storm in the SEB, or a very fast-moving plume on the NTB jet (on the south edge of the belt), or more extended disturbances in the NEB. However, recent PRO and AM studies have converged to reveal that each of these outbreaks has a similar convective origin [30, 32]. A continuing mystery is why the SEB and NTB cycles often occur within a year of each other, along with striking coloration in the equatorial zone, comprising a 'global upheaval' [13, 17].

Vortices, most prominently anticyclonic ovals, are also followed from their generation through evolution in shape, colour and speed. The largest of these is the Great Red Spot (GRS), which has been observed by amateurs since 1831 if not earlier [13]. Over that period it has shrunk by 50% of its original size (Fig. 4B), has interacted with numerous other features and spots, and has changed its drift rate. Furthermore, in recent decades its internal rotation period has shortened [18] (Fig. 4C). Nevertheless it has consistently maintained an oscillation in longitude with a 90-day period (Fig. 4A) [18,34,35]. Other long-lived white ovals include a well-defined series in the South South Temperate Zone (SSTZ), whose changing motion during 2007-2008 can be seen in Fig. 3. Recently, JUPOS analysis has revealed that anticyclonic ovals in the North North Temperate Zone (NNTZ) can also be tracked over several years, in spite of large changes in drift rate, and that one is a very long-lived Little Red Spot [36]. In the South Temperate Zone (STZ), amateurs monitored long-lived white ovals that merged in 1998-2000 to form the single large oval BA, and reported that BA turned red in 2006; this discovery by Christopher Go triggered much professional interest [36, 37]. JUPOS data have revealed how its motion is affected by South Temperate Belt (STB) segments impinging on it [15]. Amateurs have likewise tracked a long-lived white oval in the North Tropical Zone, 'White Spot Z (WSZ)' (Fig. 7, [39]) which also merged with another oval in 2013 (Fig. 8, [40]), and subsequently turned pale reddish as well [39]. These interactions of anticyclonic vortices are providing rare probes of the atmospheric dynamics and of the chemistry which generates red color on Jupiter.

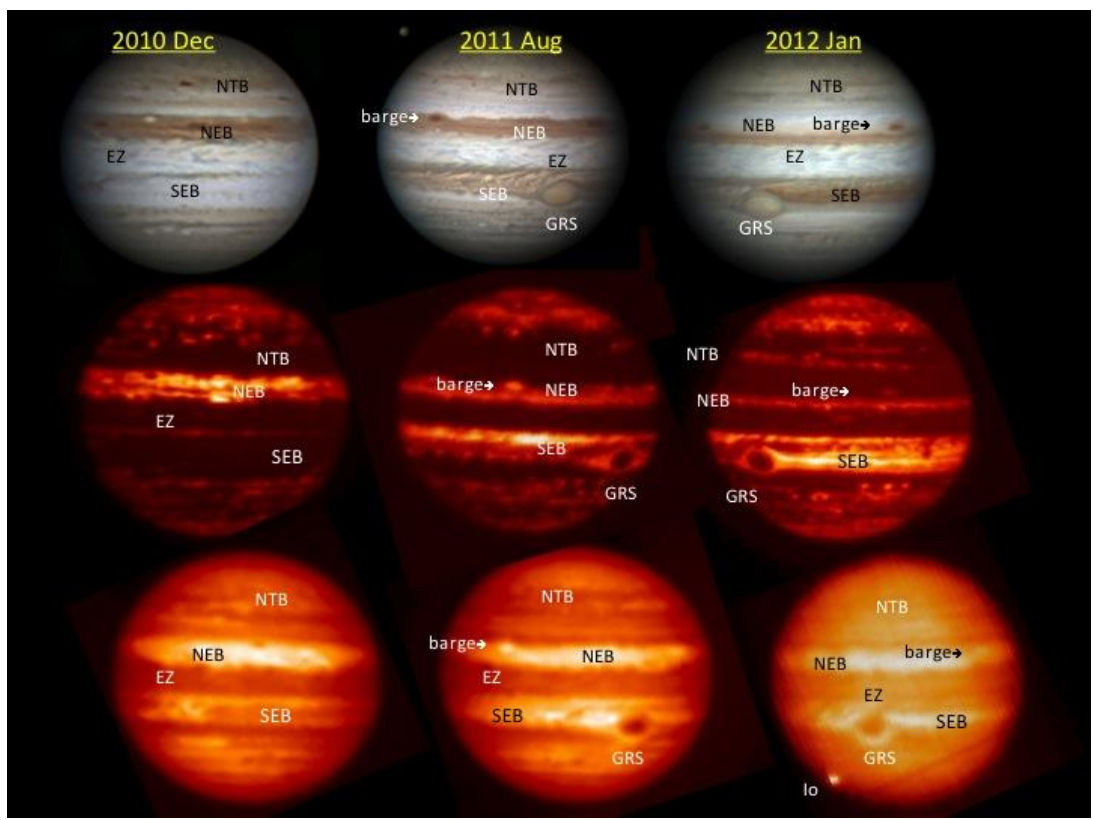

*Figure 6. Amateur RGB images (top) and professional (middle and bottom) thermal-infrared images during phases of the SEB revival and subsequent NEB fade. The middle images are taken with a filter centered at 4.78 microns, in the '5-micron spectral window', showing thermal radiation from deep levels with clouds silhouetted dark against it, sensitive to cloud all the way down to the 2-3 bar pressure levels. The lower images are at 8.7 microns, in an atmospheric spectral window between methane emission and ammonia absorption lines that is not quite so transparent and only sees down to the ~1 bar level; so it may be sensitive only to clouds at the ammonia condensation level, and nothing deeper. (All images with North up).*

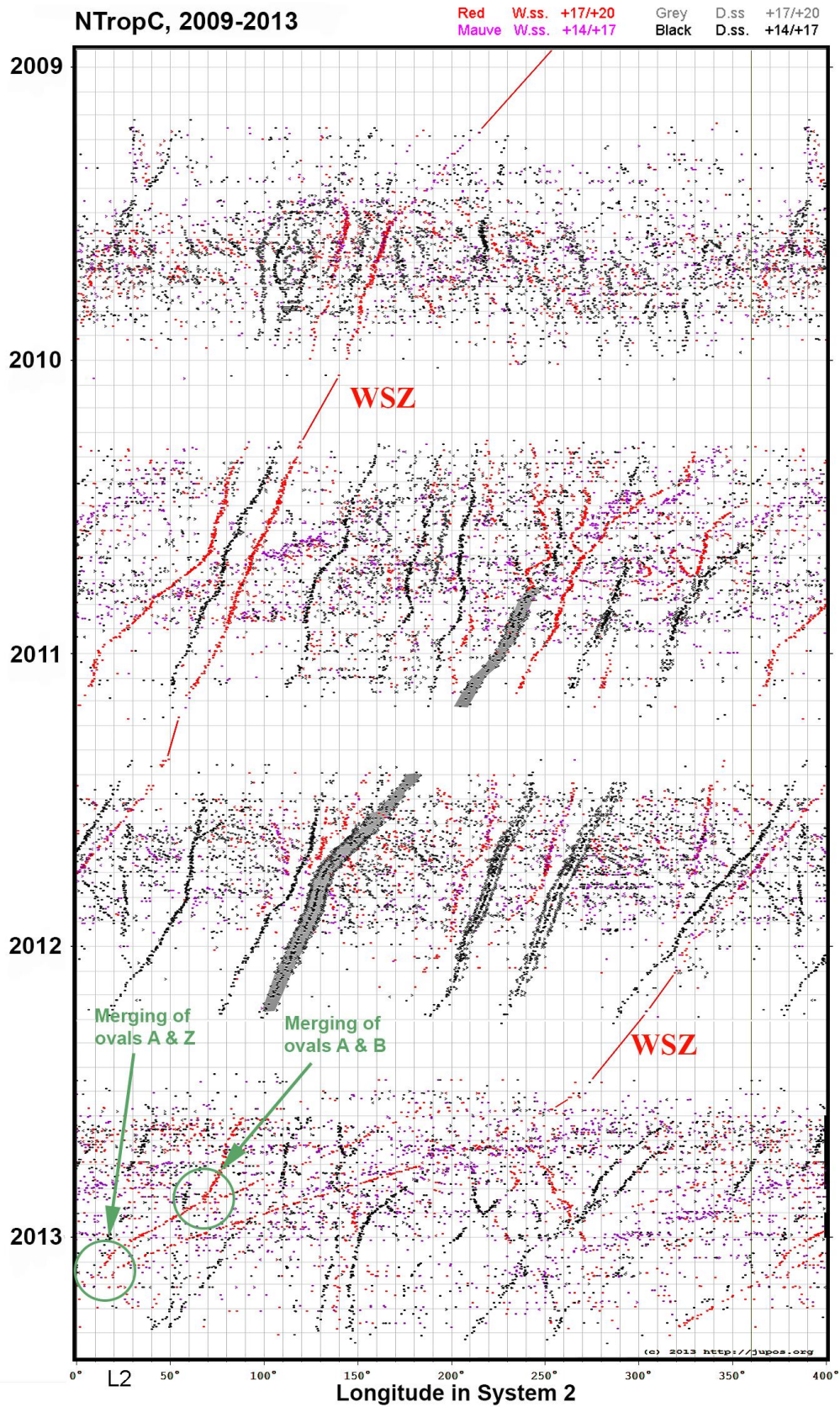

*Figure 7. A JUPOS drift chart for North Tropical domain, for 2009-2013 (in L2, A latitude key is given at the top). The track of White Spot Z (WSZ) is marked in red. The merging of ovals A and B is marked (upper right green circle). The merged oval A/B (referred as A) a few weeks later merged with WSZ (lower left green circle). We can also see the tracking of other persistent bright and dark ovals. (Adapted from [39]).*

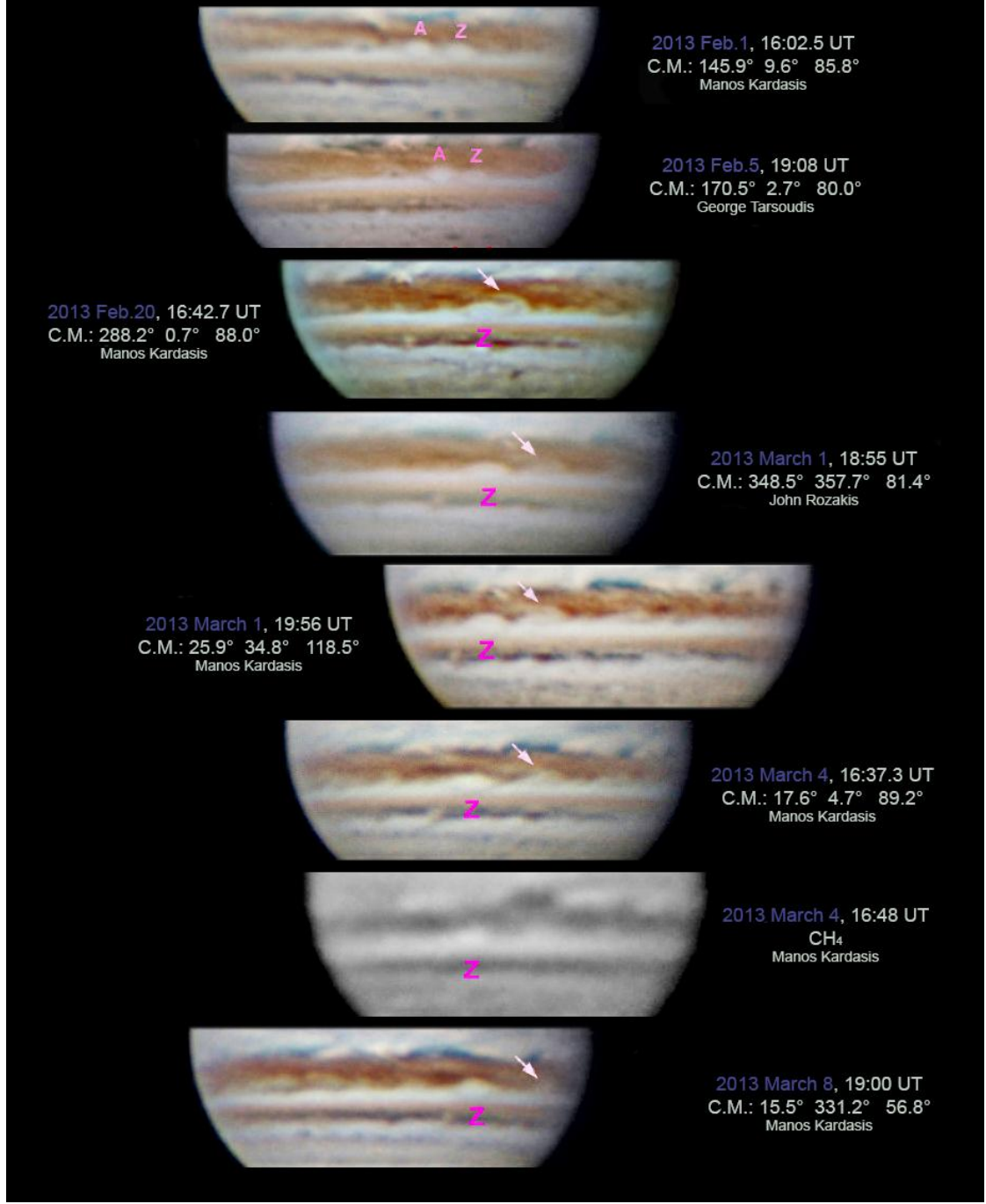

***Figure 8.*** *Interaction of NEBn ovals A&Z at 16° North in early 2013 (the Greek observations as part of a compilation in a BAA Jupiter Section report [40], South is up.) WS-A and WS-Z interacted between February 10-19. As in previous such encounters WS-A squeezed round the South side of WS-Z and was pulled into a loop; part of it probably merged with WS-Z, while the other part moved to the following side for more than two weeks (pink arrow).*

## 2.2.2 Evolution of the atmosphere of Saturn

For Saturn, like Jupiter, much of the historical record of the atmospheric cycles has come from amateurs, including progressive variations of the belt patterns around the saturnian year (Fig. 2, bottom), and the rare 'Great White Spot' (GWS) storms which have been observed roughly once per saturnian year [41]. The GWS of 1990 was discovered by amateurs and soon observed by professionals and by the *HST* [42, 43].

Since the *Cassini* spacecraft arrival in 2004, a fruitful PRO-AM collaboration on the planet's atmosphere study has developed. The spacecraft's Radio and Plasma Wave Science instrument (RPWS) regularly observes Saturn Electrostatic Discharges (SEDs), radio signatures of lightning. As these observations lack resolution, amateurs' images at visible wavelengths helped to locate the white spots (thunderstorms) that are the sources of these SEDs [44] (Fig. 9). Furthermore, given the increasing quality and coverage of these images over the years, amateurs through their own analysis can calculate the drift rates and follow the morphological evolution of the visible white spots, even the smaller ones [45].

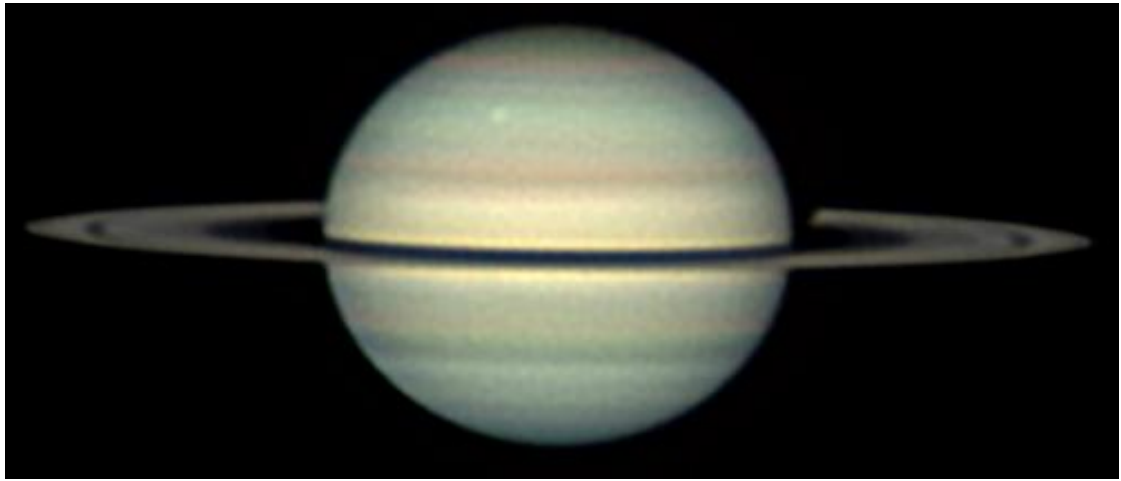

*Figure 9.* Colour image taken by Stefan Buda showing a bright storm, preceding the Central Meridian (2009 April.12, 11:02UT, South is up).

When the new Great White Spot (GWS) erupted in 2010 December [46, 47], it was independently discovered by the *Cassini* RPWS team (Dec. 5) and by the amateur Sadegh Ghomizadeh (Iran, Dec. 8); its first appearance was then identified on images taken on Dec. 5 by *Cassini* and by the amateur Toshihiko Ikemura (Japan). *Cassini* was then instructed to take high-resolution multispectral images starting several weeks later, while amateurs monitored the storm's phenomenal growth in the meantime. Hence, amateurs could fully participate in the scientific study of this rare and major event (see Fig. 10). They were fully involved in the studies from the beginning of this storm till its end [46-51]. Their work on the analysis of approximately 100 spots contributed to the study of Saturn's wind profile over the latitude range of the GWS [47].

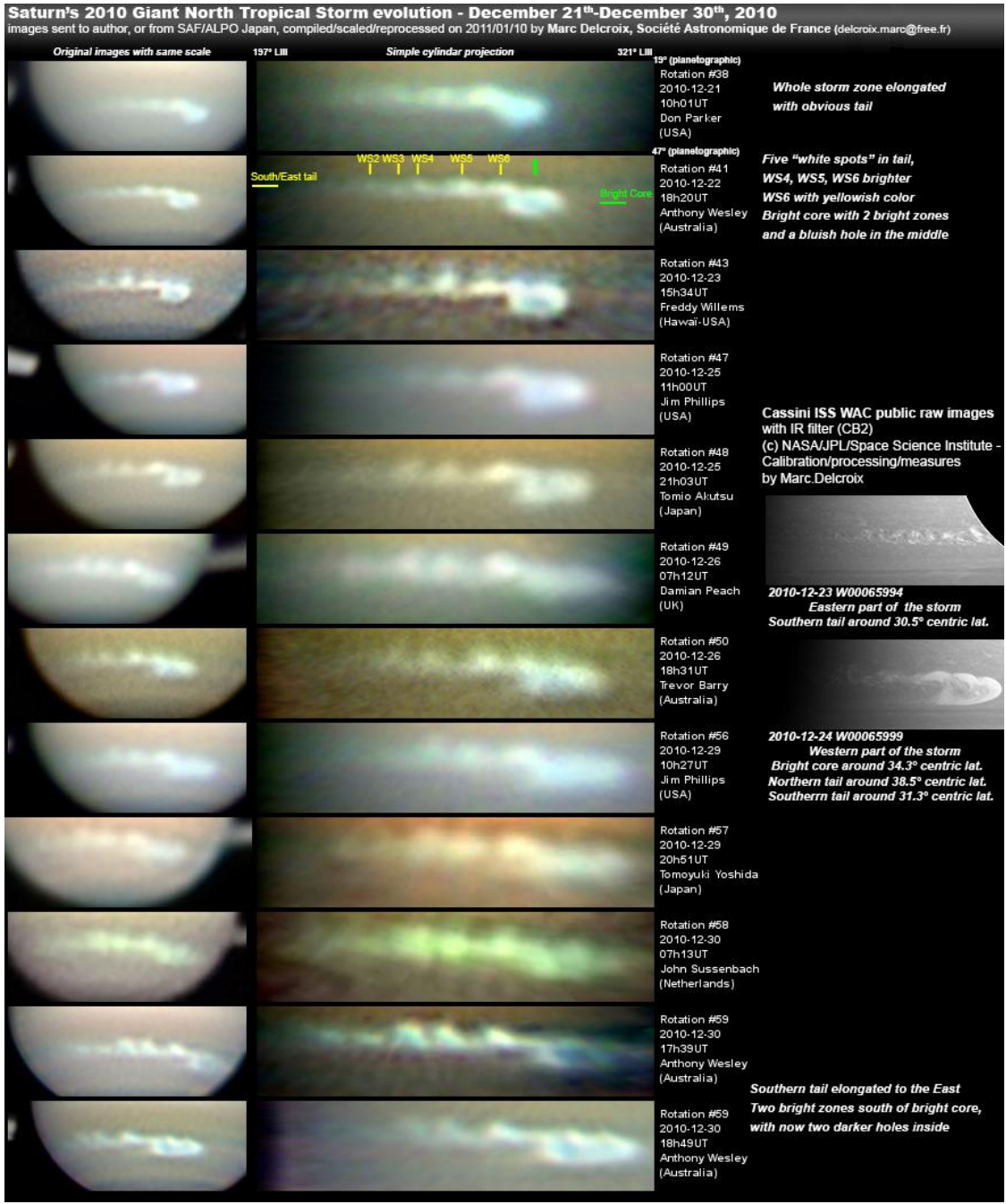

*Figure 10.a:* *Evolution of Saturn's growing Great White Spot in 2010-11 from amateur images in scale (left), maps made from the amateur images (center) and images by Cassini spacecraft (right).*

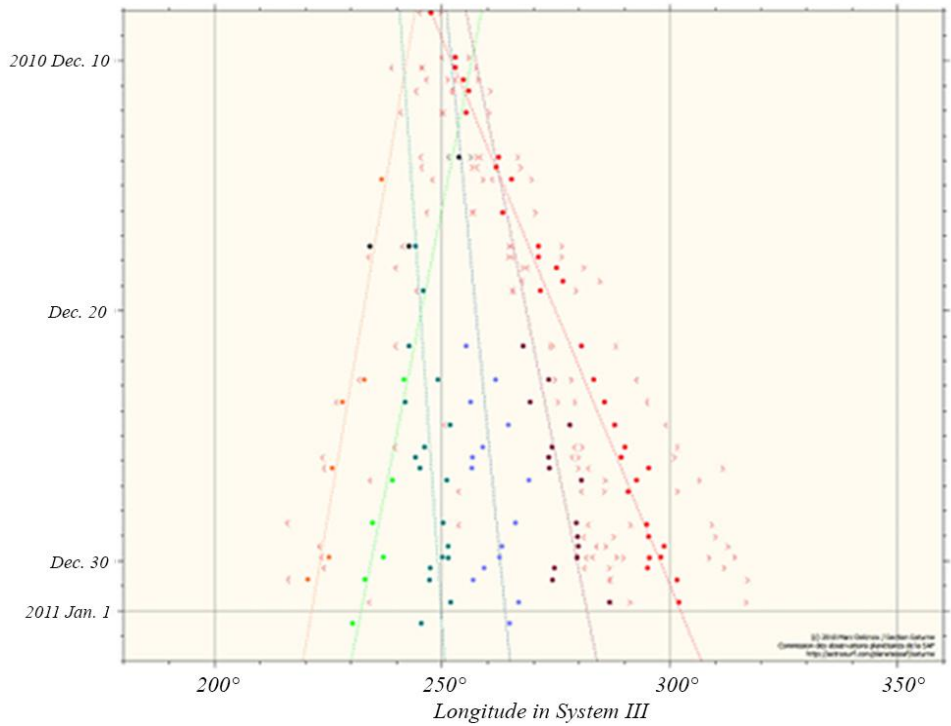

*Figure 10.b:* *Tracking of white spots. Every dot is a position measurement of a spot used to track in detail their motion and drift rate (speed). Analysis by co-author M. Delcroix from selected amateur and Cassini images (South is up).*

As the storm decayed, a large dark vortex appeared in the same latitude band, which attracted professional interest [50]. This has been imaged by the *Cassini* spacecraft and the results are presented at [52]. It has also been imaged and tracked by amateurs (Fig. 11). This spot is one of the longest-lived features ever recorded on Saturn.

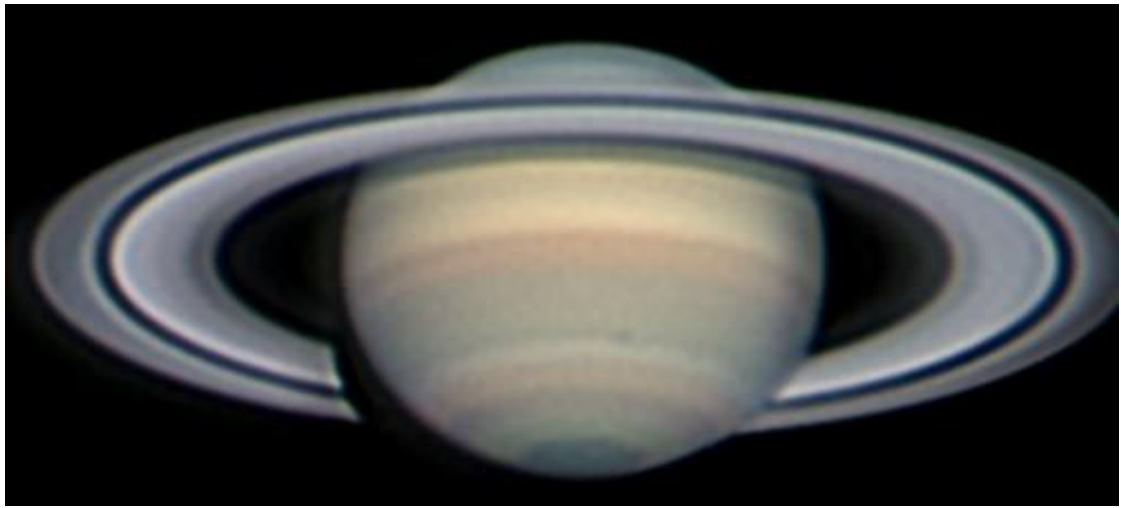

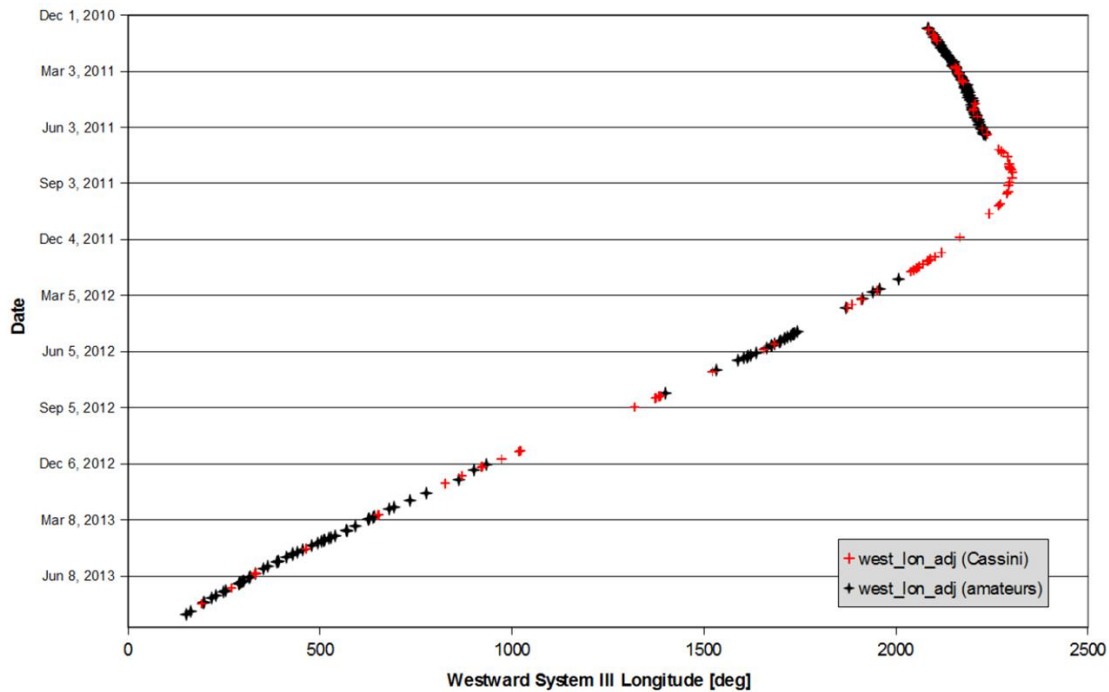

*Figure 11. Top: Colour image of Saturn by Trevor Barry showing the dark northern hemisphere vortex approaching the Central Meridian (2013 Feb 15, 18:11 UT, South is up), Bottom: Tracking of GWS dark oval longitude since the 2010 GWS, complementary covered by amateur data (black) and Cassini data (red)[52].*

In 1980-1 the Voyager spacecraft observed for the first time a strange, hexagon-shaped structure in the upper clouds of Saturn's north pole [54]. The hexagon seemed to have almost-stationary motion in the System III rotation period. More recently, amateurs using the latest imaging techniques have been able to detect the hexagon (Fig. 12). A PRO-AM collaboration with detailed observations from various geographical locations confirmed the hexagon configuration and determined that its movement remains extremely stable and may be linked with deeper layers of the planet [55], although its movement is non-zero relative to Saturn's System III longitudes. Such observations made over a long period of time will allow the rotation period of the hexagon to be established plus any long term changes identified, even after the *Cassini* mission ends.

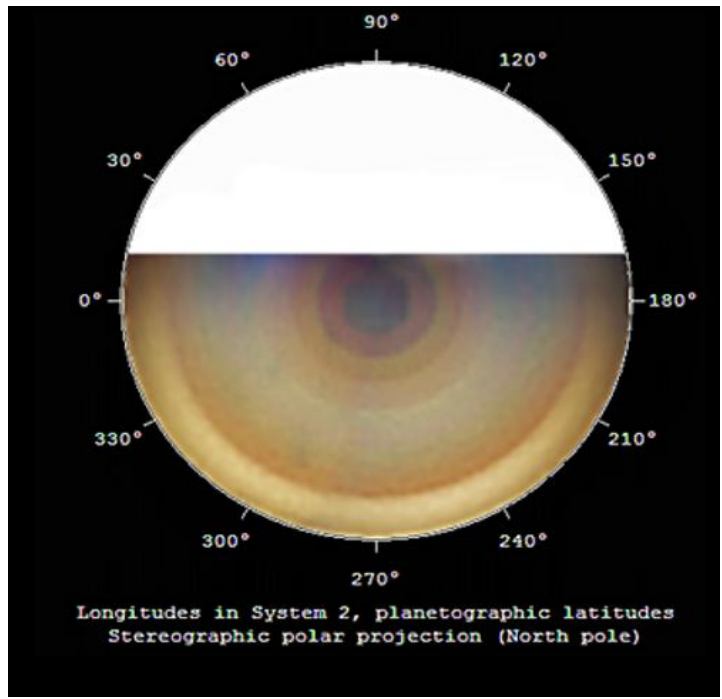

*Figure 12. North Polar projection map of Saturn showing the North polar hexagon. This map was generated by Darryl Milika and Pat Nicholas using WinJUPOS, from their RGB images taken on February 16, 2014.*

## 2.3 Ground-based space mission support

The importance of PRO-AM collaborations is also evident in the support of spacecraft missions, like the the *Galileo* and *Cassini* outer-planet missions to Jupiter and Saturn, respectively. Galileo suffered from a failure of its high-gain antenna to deploy properly and its science team had to contend with an extremely reduced data rate. Galileo's Atmospheric Working Group decided to develop the best science via "campaigns" in which specific features would be targeted and viewed by all the remote-sensing instruments at the same time, from the ultraviolet through the thermal infrared. Although some targets were belts or zones, not confined in longitude, most were discrete features. Pointing for these campaigns needed to be developed months ahead of time and loaded into the spacecraft memory and it was important to understand the drift rate of the features in order to make good predictions of their location at the scheduled time of the observations. Although some of this work was aided by a program that routinely tracked features using a suite of near-infrared wavelengths from NASA's IRTF atop Mauna Kea, Hawaii, other features were not clear in the near infrared. For that purpose, the mission relied on observations from amateurs. Amateur observations were also used to verify the results from the NASA IRTF program [56].

The operations of the *Cassini* spacecraft were not so difficult and routine observations were and are still being made by many instruments. Nevertheless, amateur observations of Saturn have played key roles. Because *Cassini's* remote-sensing instruments cannot point everywhere at the same time, it falls to the amateur community to inform the *Cassini* science team about rapidly evolving features in Saturn's atmosphere. A very good example, already described in section 2.2.2, occurred at the very beginning of the great northern hemisphere storm of 2010-2011. In that case *Cassini*'s detection of an intense radio burst coincided with amateur detection of a new bright cloud, which marked the initial upwelling of this giant thunderstorm. Remnants of that storm are still being tracked by both professional and amateur (see section 2.2.2).

The science team for the *Juno* mission will be soliciting Earth-based observations during its remote-sensing orbits in 2016-2017 to provide contextual spatial information to supplement its narrow coverage of the planet in each orbit, as well as temporal coverage that is relevant to the evolution of the features that will be observed [57]. Co-author G. Orton will be serving as the point of contact. A similar involvement of amateurs is foreseen for the European Space Agency's JUICE mission in the time frame of the 2030s [58].

## 2.4 Investigation of impacts on Jupiter's atmosphere

Since the 1994 impacts of the fragments of comet Shoemaker-Levy 9, comprehensively observed by both professionals and amateurs, amateur observations of Jupiter have resulted in the discovery of the dark visible trace of an impact in 2009, and three fireballs in its atmosphere produced by the impacts of smaller objects. The impact cloud discovered by Anthony Wesley on July 19, 2009, was immediately followed up by professionals using infrared observatories and *HST*. The fireballs were detected on June 3, 2010, August 20, 2010, and September 10, 2012, and the first of these was also followed up by professionals using infrared observatories and *HST* [59], although no residual trace of the impact could be detected. While the size of the object which left a trace in 2009 is estimated to have been between 200 and 500m, the later fireballs were in the category of 5-20 m in size, comparable to the recent Chelyabinsk airburst [60].

There could be more impacts still undetected on stored amateur videos. Hence, a project was developed called DeTeCt with software and database to search for fireballs on existing videos (Fig. 13), in order to constrain the rate of detectable Jovian impacts [61, 62]. As of the writing of this paper (January 2015), more than 25.4 days total time of videos (from 25758 videos from 41 observers acquired between February 2004 and January 2015) were analyzed. The lack of positive detections in this preliminary data set merely suggests that the observable Jovian impact rate is no more than 15 per year. Further data will provide a more accurate impact rate, which will be valuable in various fields of research (e.g. dating the surfaces of the Galilean moons, and better estimation of the small-body population in the outer solar system). Consequently, there is no doubt that PRO-AM collaborations contribute actively to the scientific study of the impacts on Jupiter [59], [61].

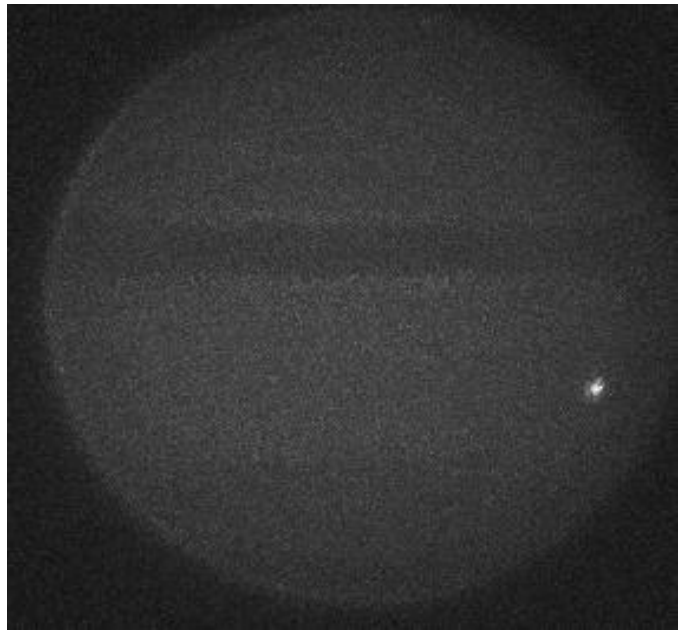

*Figure 13.* *A detection image generated by DeTeCt software (image with maximum value minus mean value for each aligned pixel) from an acquisition from Christopher Go on June 3,2010,of a real detected impact [60] (South is up).*

## 2.5 Stellar occultations as atmospheric probes

The structure and variability of the upper atmospheres of the giant planets may be investigated by occultation techniques [63]. When a spacecraft is near the planet, its radio signal may be monitored from the ground as it disappears and reappears behind the planet [64,65]. The spacecraft itself may be used as an observation platform to monitor e.g. the attenuation of a star's flux in the UV due to absorption [66, 67].

In contrast, ground-based photometric monitoring of stellar occultations measure the attenuation of starlight by the planet's intervening atmosphere due to differential refraction. It is complementary to the other techniques because it samples a different pressure regime (1-100 microbars) than the other two methods (radio: 1-1000 millibars; UV: <1 microbar). This requires a sufficiently bright star to act as source and such opportunities are not frequent.

A recent example was the occultation of the bright star 45 Capricorni (V=5.5 mag) by the planet Jupiter on the night of 3-4 August, 2009. Its occurrence during the summer motivated a PRO-AM campaign to observe it [68, 69]. In this sense, this undertaking was different than previous campaigns [70-77] in that it featured strong participation by amateur astronomers equipped with 0.4m-class instruments. To minimize the contribution of Jupiter's atmosphere to the stellar flux, observations took advantage of deep absorption bands in Jupiter's spectrum in the 0.89-μm methane band (Fig. 14).

The occultation was successfully recorded from 9 different locations in Namibia, Greece, Brazil and Spain. Full details of the observations, data reduction and scientific analysis may be found in [69]. Here we focus only on the main findings.

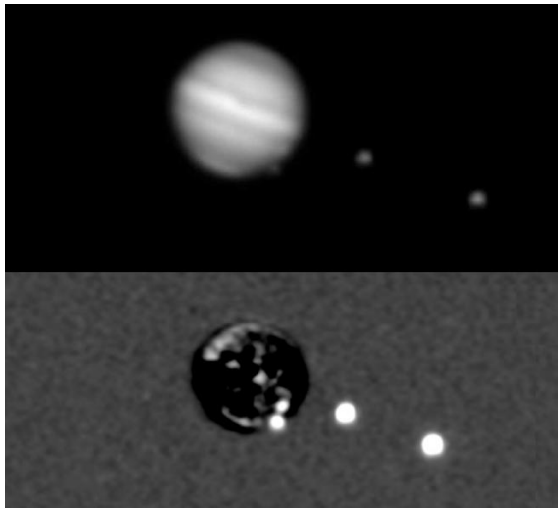

*Figure 14. Sample frame of the occultation of 45 Cap by Jupiter on 3/4, August, 2009. The frame is from a series obtained with a 0.5m instrument at Sabadell observatory in Spain (Minor Planet Center observatory code 619) in the 0.89 μm methane band before (top) and after (bottom) subtraction of a template of Jupiter. The star 45 Cap can be seen impinging on Jupiter's disk near the 5 o'clock position. The bright sources to the right of Jupiter are the moons Io and Europa (North is Up).*

The effective scale height of the atmosphere in the region probed by the occultation was found to be in the range 20-30 km, consistent with previous estimates for the mid-latitude regions of Jupiter from past occultation measurements [72-76] (but somewhat lower than that measured in the equatorial regions [77]).

A subset of the light curves was of sufficient quality (i.e. of low enough systematic and random error) to be inverted into atmospheric temperature and density as a function of height or an equivalent parameter (e.g. pressure), following previously applied algorithms [78]. In the interval 3 – 10 microbar, the profiles appear isothermal with a temperature in the range 160 – 170 K for both ingress and egress (Fig. 15 displays ingress profiles), consistent with the calculated temperature of $164 \pm 2$ K (see Fig. 4 in [69]). They are in agreement with two profiles [75] obtained at the same planetographic latitude as our measurements (albeit in the northern hemisphere of the planet) and compare favorably with temperatures of 155 – 165 K measured in situ by the Atmospheric Structure Investigation instrument suite on board the *Galileo* Entry Probe over the same pressure range in Jupiter's equatorial region [79].

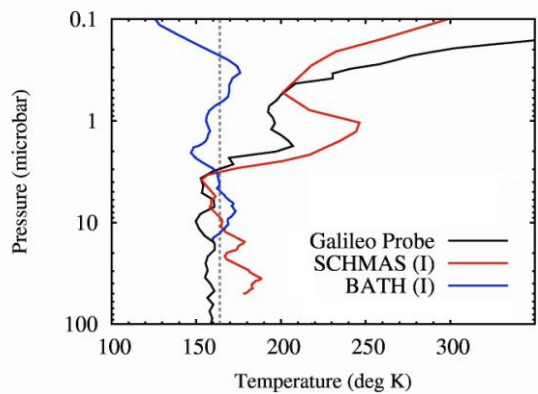

*Figure 15.* *Atmospheric profiles at ingress compared with in situ data from the Galileo Probe ASI investigation (black curve). The dotted vertical line indicates a calculated weighted mean value of the isothermal solutions obtained from the analysis (see [67] for details).*
*Abbreviations are as follows: BATH: 0.5m, Hakos, Namibia; SCHMAS: 1.52m, Teide, Canary Ils.*

A number of non-isothermal, statistically significant, features (alluded to in [75]) was also identified (see Fig. 7 in [69] and discussion therein). They display different distributions at ingress and engress, which has been observed previously (see Fig. 10 in [75]). Current data sampled similar planetographic latitudes between ingress (49 – 55 South) and egress (50 – 58 South) while in [75] the latitudes sampled were 73 North and 70 North at ingress and egress, respectively. One interpretation of these differences is either longitudinal or temporal variability at this barometric altitude range in Jupiter's atmosphere. Simultaneous photometric monitoring of future occultation events from as many geographically dispersed sites as possible will help resolve this.

In the next decade, there will be additional opportunities to apply this powerful method to investigate Jupiter's upper atmosphere and look for spatial/temporal variability. The events on April 12, 2016, and on April 2, 2021, merit special attention as they both involve bright stars (<8 mag). The 2016 event, visible from areas of Europe, Africa and the Middle East, samples similar planetographic latitudes at Jupiter as the 45 Cap event [80]. It may take on added significance in view of the expected arrival of the *Juno* mission to the Jovian system [81]. The 2021 event is visible from North and South America and samples near-equatorial Jovian latitudes. In both cases, the star is also occulted by one of the Galilean satellites, Ganymede in 2016 and Io in 2021. These secondary events may be used to constrain the existence of tenuous atmospheres around these satellites.

## 3. Summary and Conclusions

Interactions between amateur and professional astronomers have changed significantly in the digital era, from an occasional exchanging of individual images to a sustained collaboration through coordinated global networks of amateur astronomers. Today, amateur astronomers with sophisticated equipment and software provide a number of valuable resources to the professional astronomers: a large source of labour, or extension of the professional astronomer's group; a vast collection of data that provides both legacy and temporal information and finally, as ambassadors of science, a community that can help to build bridges between the scientific community and the general public. From the professional astronomer/scientist's perspective, given the vast amounts of data acquired through various projects, the natural progression to interactive collaborations between these two communities is tremendously beneficial for science and outreach. The synergy between the two communities has changed the paradigm for scientific research, with the amateur community identifying changes in the atmospheres of giant planets and discovering new phenomena, which allows the professional community to maximize the available resources for scientific results. Specifically, in the case of Jupiter and Saturn, the integration of amateur and professional data has now identified timescales for various types of intrinsic changes, for example allowing a very detailed characterization of Jupiter's 2010 SEB revival and Saturn's 2010-11 GWS storm. Extrinsic changes caused by various types of impacts have allowed the creation of a sub-discipline for observations and models. Moreover, PRO-AM studies of rare occultation events allow an investigation of the temperature, pressure, and density profiles in the upper atmospheres of giant planets.

Therefore, the continuous drive by the amateur community to push the limits and the corresponding value of their observations, as presented in this paper, has resulted in a continuous growing number of collaborations with the professional community. We strongly believe that the synergy between the two communities offers an invaluable resource in the advance of our understanding of Jupiter and Saturn.

## Acknowledgments

The authors acknowledge the valuable comments provided by the referees Richard McKim and Leigh Fletcher. E. Kardasis and J.H. Rogers acknowledge the work of JUPOS team. E. Kardasis thanks especially Grischa Hahn and Marco Vedovato from JUPOS team for their help, and his wife Dimitra Kotsa for her continuous support. The authors are grateful also to all observers around the world, as without their contributions the current paper would not materialize. This research has made use of NASA's Astrophysics Data System.

## References

[1] O. Mousis *et al.*, 'Instrumental methods for professional and amateur collaborations in planetary astronomy', *Experimental Astronomy,* **38**(1-2), 91-191, (2014)
[2] D. Boyd, 'Pro-am collaboration in astronomy – past, present and future', *J. Brit. Astron. Assoc.*, **121**(2), 73-90, (2011), (Presidential address)
[3] British Astronomical Association (BAA), website: **http://britastro.org**
[4] Association of Lunar and Planetary Observers (ALPO), website: **http://alpo-astronomy.org**
[5] Association of Lunar and Planetary Observers in Japan (ALPO Japan), website: **http://alpo-j.asahikawa-med.ac.jp/indexE.htm**
[6] Société Astronomique de France (SAF), website: **http://www2.saf-lastronomie.com**
[7] Registax, Image processing software, website: **http://www.astronomie.be/registax**

[8]   Autostakkert, Image processing software, website: **http://www.autostakkert.com**
[9]   IOPW-PVOL database, website: **http://www.pvol.ehu.es**
[10]  Database for Object Positions on Jupiter (JUPOS), website: **http//www.jupos.org**
[11]  J. H. Rogers, ' Multispectral imaging of the EZ and NTB coloration events', Report 2012-4 of Brit. Astron. Assoc. - Jupiter section, (2012 Sep. 20), available at:
      **http://www.britastro.org/jupiter/2012_13report04.htm**
[12]  B. M. Peek, 'The Planet Jupiter', *Faber & Faber,* London, (1958)
[13]  J. H. Rogers, 'The Giant Planet Jupiter', *Cambridge Univ.,* (1995)
[14]  G. Hahn, M. Jacquesson, 'Jupiter : Longitudinal drifts computation from image pairs',  available at:
      **http://www.grischa-hahn.homepage.t-online.de/astro/winjupos/LongDrifts/Jupiter_LongDriftDetermination_English.pdf**
[15]  J. H. Rogers, *et al.*, 'Jupiter's South Temperate   domain: Behaviour of long-lived features and jets, 2001-2012', Brit. Astron. Assoc. - Jupiter section, (2013 Jul 16), available at:
      **http://www.britastro.org/jupiter/stemp2013.htm**
[16]  N. Barrado-Izagirre, 'Jupiter's zonal winds and their variability studied with small-size telescopes', *Astronomy & Astrophysics,* **554** (A74), (2013)
[17]  British Astronomical Association,  'Jupiter section', website:
      **http://www.britastro.org/jupiter/section_reports.htm** and
      **http://www.britastro.org/jupiter/reference.htm**
[18]  J. H. Rogers, 'The accelerating circulation of Jupiter's Great Red Spot',   *J.  Brit.  Astron. Assoc.,* **118** (1), 14-20, (2008), and ibid. (2012), available at:  **http://www.britastro.org/jupiter/2012_13report07.htm.**
[19]  A. A. Simon *et al.*, 'Dramatic Change In Jupiter's Great Red Spot From Spacecraft Observations' *Astrophysical Journal Letters,* **797**(2), (2014)
[20]  E. Kardasis, 'Jupiter 2014/03/14-15 strip maps & Methane 889nm band report',  available at:
      **http://www.hellas-astro.gr/articles/astromanos-2014-04-01-1447**
[21]  L. N. Fletcher *et al.*, 'Jovian temperature and cloud variability during the 2009-2010 fade of the South Equatorial Belt', *Icarus,* **213**, 564-580, (2011)
[22]  Grupo de Ciencias Planetarias (Group of planetary scientists), University of the Basque Country in Bilbao (Spain), website: **http://www.ajax.ehu.es**
[23]  A. Sánchez-Lavega *et al.*, 'A disturbance in Jupiter's high-speed north temperate jet during 1990', *Icarus,* **94**, 92–97, (1991)
[24]  A. Sánchez-Lavega *et al.*, 'The South Equatorial Belt of Jupiter, II: The Onset and Development of the 1993 Disturbance', *Icarus,* **121**, 18–29, (1996)
[25]  G. S. Orton *et al*., 'Spatial organization and time dependence of Jupiter's tropospheric temperatures, 1980-1993', *Science* **265**, 625-631, (1994)
[26]  J. H. Rogers, T. Akutsu, G.S. Orton, 'Jupiter in 2000/2001: Part II: Infrared and ultraviolet wavelengths: A review of multispectral imaging of the Jovian atmosphere', *J. Brit. Astron. Assoc.,* **114**, 313–330, (2004)
[27]  J. H. Rogers, H-J. Mettig, and D. Peach, 'Renewed acceleration of the 24°N jet on Jupiter',  *Icarus,* **184**, 452-459, (2006)
[28]  A. A. Simon-Miller *et al.*, 'Longitudinal variation and waves in Jupiter's south equatorial wind jet', *Icarus,* **218,** 817–830, (2012)
[29]  J. H. Rogers *et al.*, 'Jupiter in 2000/2001 Part III: The South Equatorial Disturbance: A large-scale wave in a prograde jet',  *J. Brit. Astron. Assoc.,* **115**(2), 70-78, (2005)
[30]  A. Sanchez-Lavega, J. M. Gomez, 'The South Eq. Belt of Jupiter, I: Its Life Cycle', *Icarus*, **121,** 1–17, (1996)
[31]  J. H. Rogers *et al.*, 'Jupiter's North Equatorial Belt: An historic change in cyclic behaviour with acceleration of the North Equatorial jet',  *European Planetary Science Congress 2013*, **8**(384), London
[32]  A. Sanchez-Lavega *et al.*, 'Depth of a strong jovian jet from a planetary-scale disturbance driven by storms', *Nature,* **451**, 437–440, (2008)
[33]   R. S. Giles *et al.*, 'The 2010-2011 revival of Jupiter's South Equatorial Belt.',  *European Planetary Science Congress 2013*, **8**(33), London
[34]  J. M. Trigo-Rodriguez *et al.*, 'The 90-day oscillations of Jupiter's Great Red Spot revisited', *Planetary and Space Science,* **48**, 331–339, (2000)
[35]  G. Hahn, 'The 90-day oscillation of the Jovian Great Red Spot',  *J. Brit. Astron. Assoc.,* **106,** 40-43, (1996)
[36]  J. H. Rogers, G. Adamoli & H-J Mettig, 'Jupiter's high-latitude storms: A Little Red Spot tracked through a jovian year', *J. Brit. Astron. Assoc.,* **121**(1), 19-29, (2011)
[37]  E. Garcia-Melendo *et al.*, 'The jovian anticyclone BA: I. Motions and interaction with the GRS from observations and non-linear simulations', *Icarus,* **203**, 486-498, (2009)
[38]  M. H. Wong *et al.*, 'Vertical structure of Jupiter's oval BA before and after it reddened: What changed?', *Icarus,* **215**, 211-225, (2011)
[39]  J. H. Rogers, 'White spot Z: its history and characteristics 1997-2013', (2013 Dec 18), available at:
      **http://www.britastro.org/jupiter/2013_14report03.htm**
[40]  J. H. Rogers, 'Jupiter update: NNTBs jet;  NEBn white ovals; new SED; STB near oval BA', (2013 March 18), available at: **http://www.britastro.org/jupiter/2012_13report10.htm**
[41]  R. J. McKim, 'Great White Spots on Saturn: current and historical observations', *J. Brit. Astron. Assoc.,* **121**(5), 270-273, (2011)
[42]  A. Sánchez-Lavega *et al.*, 'The Great White Spot and disturbances in Saturn's equatorial atmosphere during 1990',  *Nature,* **353**, 397-401, (1991)
[43]  A. W. Heath, R. J. McKim, 'Saturn 1990: the Great White Spot', **102**(4), 210-219, (1992)
[44]  G. Fischer *et al.*, 'Overview of Saturn lightning observations', *Proceedings to "Planetary radio Emissions VII", Austrian Academy of  Sciences Press*, 135-144, (2011)

[45] M. Delcroix and G. Fischer, 'Contribution of amateurs to Saturn's storms studies', *European Planetary Science Congress 2010,* **5**(132), Rome
[46] G. Fischer *et al.*, 'A giant thunderstorm on Saturn', *Nature,* **475**, 75-77, (2011)
[47] A. Sanchez-Lavega *et al.,* 'Deep winds beneath Saturn's upper clouds from a seasonal long-lived planetary-scale storm', *Nature,* **475**, 71-73, (2011)
[48] A. Sanchez-Lavega *et al.*, 'Ground-based observations of the long-term evolution and death of Saturn's 2010 Great White Spot', *Icarus,* **220**, 561-576, (2012)
[49] L. N. Fletcher *et al.*, 'The origin and evolution of Saturn's 2011-2012, stratospheric vortex', *Icarus,* **221**, 560–586, (2012)
[50] L. N. Fletcher *et al.*, 'Thermal Structure and Dynamics of Saturn's Northern Springtime Disturbance', *Science,* **332**, 1413, (2011)
[51] J. F. Sanz-Requena *et al.*, 'Cloud structure of Saturn's 2010 storm from ground-based visual imaging', *Icarus*, **219**, 142–149, (2012)
[52] K. M. Sayanagi *et al.*, 'Dynamics of Saturn's Great Storm of 2010-2011 from Cassini ISS and RPWS', *Icarus*, **223**, 460-478, (2013)
[53] M. Delcroix *et al.*, 'Saturn northern hemisphere's atmosphere and polar hexagon in 2013', *European Planetary Science Congress 2013*, **8**(1067), London
[54] D. A.Godfrey, 'A hexagonal feature around Saturn's North Pole', *Icarus*, **76**, 335-356, (1988)
[55] A. Sánchez-Lavega *et al.*, 'The long-term steady motion of Saturn's hexagon and the stability of its enclosed jet stream under seasonal changes', *Geophysical Research Letters*, **41**, 1425-1431, (2014)
[56] G. S. Orton *et al.*, 'Characteristics of the Galileo probe entry site from Earth-based remote sensing observations', *Journal of Geophysical Research - Planets*, **103**(E10), 22791-22814, (1998)
[57] G. Orton, 'A New Mission-Supporting Era of Amateur Astronomy: The Juno Mission and the Role of Amateur Astronomers', *European Planetary Science Congress 2012*, **7**(288), Madrid
[58] L. N. Fletcher to Kardasis et al, personal communication, 2014 October 20
[59] R. Hueso *et al.,* 'First Earth-based Detection of a Superbolide on Jupiter', *The Astrophysical Journal Letters,* **721**(2), (2010)
[60] R. Hueso *et al.*, 'Impact Flux on Jupiter: From superbolides to large-scale collisions', *Astronomy & Astrophysics,* **560**(A55), (2013)
[61] M. Delcroix *et al.*, 'Jovian impact flashes detection with DeTeCt software project', *European Planetary Science Congress 2013*, **8**(812-1), London. Project page available at: **http://www.astrosurf.com/planetessaf/doc/project_detect.shtml**
[62] R. Hueso *et al.*, 'Flux of impacts in Jupiter: From superbolides to large-scale meteorites', *IAA Planetary Defense Conference 2013*
[63] J. L. Elliot, and C. B. Olkin, 'Probing Planetary Atmospheres with Stellar Occultations', *Annu. Rev. Earth Planet. Sci.*, **24**, 89-123, (1996)
[64] G. F. Lindal *et al.*, 'The Atmosphere of Jupiter: An Analysis of the Voyager Radio Occultation Measurements', *J. Geophys. Res.,* **86**, 8721, (1981)
[65] G. F. Lindal, 'The atmosphere of Neptune: An analysis of radio occultation data acquired with Voyager 2', *Astron. J.,* **103**, 967-982, (1992)
[66] M. C. Festou *et al., '*Composition and thermal profiles of the Jovian upper atmosphere determined by the Voyager ultraviolet stellar occultation experiment', *J. Geophys. Res.,* **86**, 5717, (1981)
[67] T. K. Greathouse *et al.*, 'New Horizons Alice ultraviolet observations of a stellar occultation by Jupiter's atmosphere', *Icarus*, **208**, 293, (2010)
[68] A. A. Christou *et al*, 'The 2009 occultation of the bright star 45 Cap by Jupiter', *European Planetary Science Congress 2012,* **7**(378), Madrid
[69] A. A. Christou *et al., '*The occultation of HIP 107302 by Jupiter', *Astronomy & Astroph.*, **556**(A118), (2013)
[70] E. Raynaud *et al.*, 'A re-analysis of the 1971 Beta Scorpii occultation by Jupiter: study of temperature fluctuations and detection of wave activity', *Icarus*, **168**, 324-335, (2004)
[71] W. A. Baum and A. D. Code, 'A photometric observation of the occultation of a Arietis by Jupiter', *Astron. J.,* **58**, 108-112, (1953)
[72] W. B. Hubbard *et al.*, 'The occulation of Beta Scorpii by Jupiter and Io. I. Jupiter', *Astron. J.*, **77**, 41-59, (1972)
[73] L. Vapillon, M. Combes, J. & Lecacheux, 'The β Scorpii occultation by Jupiter. II. The temperature and density profiles of the Jovian upper atmosphere', *Astron.&Astrophysics*, **29**, 135-149, (1973)
[74] J. Veverka *et al.*, 'The occultation of β Scorpii by Jupiter I. The structure of the Jovian upper atmosphere', *Astron. J.*, **79**, 73, (1974)
[75] E. Raynaud *et al.*, 'The 10 October 1999 HIP 9369 occultation by the northern polar region of Jupiter: ingress and egress lightcurves analysis', *Icarus*, **162**, 344, (2003)
[76] E. Raynaud *et al.*, 'A re-analysis of the 1971 Beta Scorpii occultation by jupiter: study of temperature fluctuations and detection of wave activity', *Icarus*, **168**, 324, (2004)
[77] W. B. Hubbard *et al.*, 'The Occultation of SAO 78505 by Jupiter', *Icarus*, **113**, 103–109, (1995)
[78] L. H. Wasserman, and J. Veverka, 'On the reduction of occultation light curves', *Icarus*, **20**, 322-345, (1973)
[79] A. K. Seiff *et al*., 'Structure of the atmosphere of Jupiter: Galileo Probe measurements', *Science*, **272**(5263), 844-845, (1996)
[80] Smithsonian Astrophysical Observatory, Occultations predictions, available at: **http://tdc-www.harvard.edu/occultations**
[81] S. J. Bolton & the Juno Science Team, 'Galileo's Medicean Moons: their impact on 400 years of discovery', C*ambridge: Cambridge University Press*, *Proc. IAU Symp.*, **269**, 92, (2010)